# Focal plane wave-front sensing algorithm for high-contrast imaging

DOU JiangPei [1, 2†], REN DeQing[1, 3], ZHU YongTian[1] & Zhang Xi [1, 2]

1. National Astronomical Observatories/Nanjing Institute of Astronomical Optics & Technology, Chinese Academy Science, Nanjing 210042
2. Graduate School of the Chinese Academy of Sciences, Beijing 100049
3. Physics & Astronomy Department, California State University Northridge, 18111 Nordhoff Street, Northridge, California 91330

**High-contrast imaging provided by a coronagraph is critical for the direction imaging of Earth-like planet orbiting its bright parent star. A major limitation for such direct imaging is the speckle noise that is induced from the wave-front error of an optical system. Based on the first-order approximation, we derived an algorithm for the wave-front measurement directly from 3 focal plane images. The simulation shows that the reconstructed wave-front is consistent with the original wave-front in the first order, which indicates that such an algorithm is a promising technique for the wave-front measurement for the high-contrast imaging.**

high-contrast imaging, multi-image algorithm, speckle nulling technique, wave-front sensing

Discovering life on another planet will be one of the most important scientific advances of this century. The search for life requires the ability to detect photons directly from planets and the use of spectroscopy to analyze physical and atmospheric conditions. The direct detection of earth-like low-mass planets is extremely challenging since the brightness ratio between an Earth-like planet and its parent star is in the order of $10^{-9}$ in the visible, and the diffraction of starlight is much stronger than the nearby planet image. A high-contrast imaging coronagraph can be used to suppress the diffraction light from the bright star, so that the nearby planet can be detected in the ideal case [1−3]. However, the actual performance of a high-contrast coronagraph is limited by the wave-front error of the coronagraphic system, which induces speckle noise [4]. Because of the large bright difference, the local speckles are much brighter than the planet image, making the direct imaging impossible. To eliminate the wave-front error induced speckle noise, the speckle nulling technique [5] is introduced by the using of a deformable mirror (DM) to create a local dark zone that can be served as a discovery area for the planet imaging.

The most important procedure for the dark zone correction is the measurement of the wave-front error. Since the optical paths of the wave-front sensor and the final focal plane are not exactly identical, the conventional Shack-Hartmann wave-front sensor will introduce the so called non-common path error, making it not suitable for the high-contrast imaging. To avoid such a non-common path error, the best approach is to reconstruct the wave-front directly from a number of images measured on the focal plane. Recently, Borde & Traub and Give'on et al. [6−7] provided different algorithms that can reconstruct the wave-front error from 3 focal plane images. Although the focal plane wave-front sensing is a promising technique, it is found that some of their assumptions for the phase reconstruction may be not proper, which may not guarantee the precision of the measured wave-front.

---
†Corresponding authour: (email: jpdou@niaot.ac.cn)

In this paper we derive our own algorithm for the wave-front reconstruction that is based on the measurement of 3 focal plane images. In Section 1, we present our algorithm in details, in which the result is totally different with that provided by Borde & Traub and Give'on et al. [6−7]. In Section 2, we provide a numerical simulation to verify the correction and precision of our algorithm. Conclusion is reached at Section 3.

## 1 Three-image focal plane measurement

As demonstrated by recent laboratory experiments of coronagraphic systems [8], an actual coronagraph can provide a high-contrast in the order of $10^{-5} \sim 10^{-6}$. Further improvement is limited by the speckle noise. Let us consider a general coronagraphic system with a circular entrance pupil. For simplicity, we assume the system operating at a monochromatic wavelength.

For an optical system with wave-front error, the electric field of the electromagnetic wave at the pupil plane for a coronagraph can be expressed,

$$E_0 = Pe^{i\Phi} \tag{1}$$

where $P$ represents the entrance pupil function of the system and it is equal to 1 inside the pupil and 0 outside the pupil (for a general telescope pupil). $\Phi$ is the phase or wave-front error that will result in spot-like speckles surrounding the bright star image on the focal plane.

The purpose of speckle nulling technique is to create a local dark zone on the focal plane that can be served as a discovery area for the planet imaging. To null out the speckles in a specific region on the focal plane, we can use a deformable mirror (DM) to create a wave-front on the pupil plane according to a specific algorithm so that the speckle noise on the focal plane is suppressed. For a coronagraph with a DM that can provide an extra wave-front by applying voltages on the DM actuators, the pupil-plane electric field can be expressed as

$$E'_0 = Pe^{i(\Phi+\Psi)} \tag{2}$$

where $\Psi$ represents the phase created by the deformation of the DM.

For such a speckle nulling approach, the most important procedure is to measure the wave-front error precisely so that the speckle noise can be cancelled efficiently. For our focal plane wave-front sensing, we measure 3 images on the focal plane that can be used to reconstruct the original wave-front error of the optical system. These images are obtained through 3 different DM configurations with phase $\Psi_k$ (k=0, 1, 2) that correspond to different DM surface shapes. For k=0, no voltage is applied to the DM actuators, which corresponds to the original wave-front error that we need to measure.

Since the starlight is much brighter than that of the planet, we can use the star images for wave-front sensing and the planet image that is much less in intensity than the star image is negligible in the wave-front sensing process without causing any significant error. The electric field of the starlight on the focal plane for a system with phase aberration and DM deformation is the Fourier transform of the aberrated electric field on the pupil plane

$$E_k = \vec{F}(Pe^{i\Phi+i\Psi_k}) = \vec{F}\{Pe^{i\Phi}[1+(e^{i\Psi_k}-1)]\} \approx \vec{F}(Pe^{i\Phi}) + \vec{F}(P(e^{i\Psi_k}-1)) \quad (3)$$

where $\vec{F}$ represents the Fourier transform of the associated function. In the above equation, we use the first-order Taylor series to calculate the $e^{i\Phi}$: in the second term, the $e^{i\Phi}$ is approximated as 1 in the case for a small $\Phi$, which results in a first-order approximation.

The associated intensity of starlight on the focal plane for each DM configuration $\Psi_k$ is square of the complex modulus of the electric field and is given as

$$I_k = \left|\vec{F}(Pe^{i\Phi}) + \vec{F}(P(e^{i\Psi_k}-1))\right|^2 \quad (4)$$

The 3 images can be recorded by a CCD camera located on the focal plane, such as the camera for the scientific observation. Let the first image, $I_0$, be taken with $\psi_0=0$ (with no voltage applied to the DM actuators and therefore no deformation). The intensity on the focal plane is then

$$I_0 = \left|\vec{F}(Pe^{i\Phi})\right|^2 \quad (5)$$

The remaining two images are taken with two DM configurations $\Psi_1$ and $\Psi_2$ that are achieved by applying different voltages to the DM actuators. Combining Eq. (4) and (5), for each image taken, we have

$$I_1 - I_0 - \left|\vec{F}(P(e^{i\Psi_1}-1))\right|^2 = 2\{\text{Re}[\vec{F}(P(e^{i\Psi_1}-1))]\text{Re}[\vec{F}(Pe^{i\Phi})] + \text{Im}[\vec{F}(P(e^{i\Psi_1}-1))]\text{Im}[\vec{F}(Pe^{i\Phi})]\} \quad (6)$$

$$I_2 - I_0 - \left|\vec{F}(P(e^{i\Psi_2}-1))\right|^2 = 2\{\text{Re}[\vec{F}(P(e^{i\Psi_2}-1))]\text{Re}[\vec{F}(Pe^{i\Phi})] + \text{Im}[\vec{F}(P(e^{i\Psi_2}-1))]\text{Im}[\vec{F}(Pe^{i\Phi})]\} \quad (7)$$

where Re and Im represent the real and imaginary parts, respectively. As one can see, only the real and imaginary parts of the $\vec{F}(Pe^{i\Phi})$ are unknown. The DM induced phases $\Psi_1$ and $\Psi_2$ can be obtained through the so called influence function according to the voltages applied on the DM actuators [9]. That is, if the DM is well calibrated, the DM induced wave-front can be calculated from the voltages applied on the actuators. The term $\vec{F}(Pe^{i\Phi})$ can be solved via eq. (6) and (7), and the existence of the solution is ensured if the following equation is not zero

$$D = \text{Re}[\vec{F}(P(e^{i\Psi_1}-1))]\text{Im}[\vec{F}(P(e^{i\Psi_2}-1))] - \text{Im}[\vec{F}(P(e^{i\Psi_1}-1))]\text{Re}[\vec{F}(P(e^{i\Psi_2}-1))] \quad (8)$$

This condition shows that to reconstruct the original phase error properly the DM deformations must be chosen carefully so that the two configurations can create two totally different electric fields for every pixel on the focal plane.

In the case of $D \neq 0$, the real and imaginary parts of the $\vec{F}(Pe^{i\Phi})$ can be derived as

$$\text{Im}[\vec{F}(Pe^{i\phi})] = \frac{(I_2 - I_0 - \left|\vec{F}(P(e^{i\Psi_2}-1))\right|^2)\text{Re}[\vec{F}(P(e^{i\Psi_1}-1))] - (I_1 - I_0 - \left|\vec{F}(P(e^{i\Psi_1}-1))\right|^2)\text{Re}[\vec{F}(P(e^{i\Psi_2}-1))]}{2D} \quad (9)$$

$$\text{Re}[\vec{F}(Pe^{i\phi})] = \frac{(I_1 - I_0 - \left|\vec{F}(P(e^{i\Psi_1}-1))\right|^2)\text{Im}[\vec{F}(P(e^{i\Psi_2}-1))] - (I_2 - I_0 - \left|\vec{F}(P(e^{i\Psi_2}-1))\right|^2)\text{Im}[\vec{F}(P(e^{i\Psi_1}-1))]}{2D} \quad (10)$$

Once the imaginary and real parts of the $\vec{F}(Pe^{i\Phi})$ are known, the aberrated electric field is a complex on the focal plane and can be reconstructed as

$$\vec{F}_R(Pe^{i\Phi}) = \text{Re}[\vec{F}(Pe^{i\Phi})]^R + i\,\text{Im}[\vec{F}(Pe^{i\Phi})]^R \quad (11)$$

The superscript $^R$ in the above equation represents the reconstructed real and imaginary parts, respectively. The reconstructed phase error $\Phi_R$ on the pupil plane is eventually obtained as

$$\Phi_R = \mathrm{Im}[\log\{\vec{F}^{-1}[\vec{F}_R(Pe^{i\Phi})]/P\}] \tag{12}$$

where $\vec{F}^{-1}$ means inverse Fourier transform of the associated functions. By performing the inverse Fourier transform and the associated logarithm operation, the phase error is then included in the imaginary part of the complex.

## 2  Numerical simulation

The correction and precision of the focal plane wave-front sensing algorithm that we discussed in Section 1 can be verified by numerical simulation. For demonstration purpose, a coma with an rms wave-front error of 0.00208 radians is used as the original wave-front. To find the original wave-front error, we perform the associated two-dimension Fourier transform between the distorted electric field in pupil plane and the image plane according to the equations discussed in Section 1.

Table 1 shows the associated parameters for the simulation. For an optimized reconstruction, the wave-fronts induced by the DM should be in the same order with that of the original wave-front error. In this simulation, we created DM induced phases in the same order as that of the original wave-front error.

**Table 1** Simulation parameters and results

|  | RMS (radians) | Strehl Ratio |
|---|---|---|
| Original wave-front error (Coma) | 0.00208 | 0.9998294 |
| DM configuration | 0.00288 | 0.9996730 |
| Reconstructed (without averaging) | 0.00211 | 0.9998247 |
| Reconstructed (with 25 wave-fronts averaged) | 0.00209 | 0.9998282 |
|  | Error ||
| Remaining error (without averaging) | 1.35% ||
| Remaining error(with 25 wave-fronts averaged) | 0.34% ||

Figure 1 shows the starlight images with different DM configurations. As we can see these point source images are distorted by DM induced wave-fronts in different configurations. The two DM configurations introduced two uncorrelated wave-fronts that ensure eq. (8) is not zero.

Figure 2 shows the original phase map to be measured and the associated phase map reconstructed by using our focal plane wave-front sensing algorithm. It clearly indicates that the reconstructed wave-front is consistent with the original one in the first order. As we can see that the first-order approximation and the reconstruction processing create some random noise that will affect the reconstruction precision. Compared with original wave-front to be measured, the direct reconstructed wave-front algorithm delivers accuracy with 1.35% remaining error; however, such a randomly changed noise can be further reduced by averaging a number of reconstructed wave-fronts in which the DM configurations are chosen randomly. Figure 3 shows an improved result by averaging 25 re-constructed wave-front maps, in which the remaining wave-front error is reduced to 0.34%.

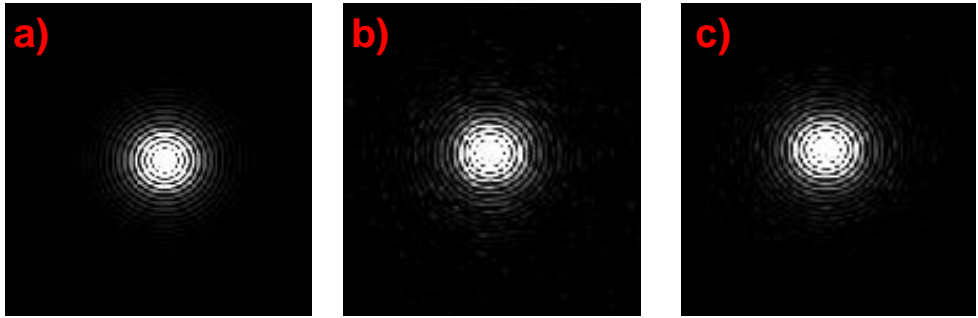

**Figure 1** a) Distorted starlight image of the system with original wave-front error; b) Starlight image with DM Configuration 1; c) Starlight image with DM Configuration 2.

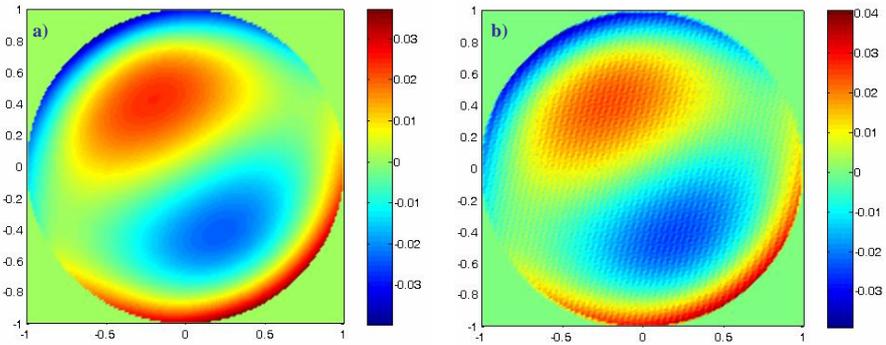

**Figure 2** a): Original wave-front map; b): Reconstructed wave-front map (without averaging)

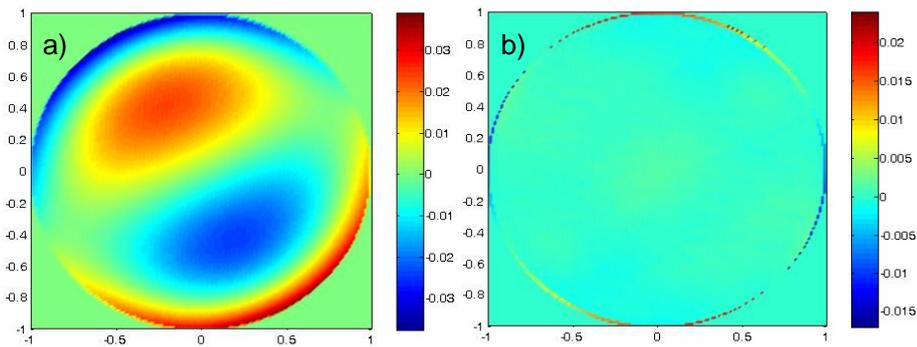

**Figure 3** a): Reconstructed wave-front map (averaged); b): Remaining error by subtracting the reconstruct wave-front with the original wave-front.

We also used other wave-front maps to replace the coma for simulations, and achieved the same result, which gives us confidence that the performance of the focal plane wave-front sensing is reliable and suitable for the wave-front measurement for the high-contrast imaging.

## 3    Conclusion

Using three images on the focal plane, wave-front error can be measured and reconstructed with high precision, which is demonstrated by our numerical simulation. The focal plane wave-front sensing algorithm we discussed can eliminate the so called non-common path error and is suitable for the wave-front measurement for the high-contrast imaging. Although we have only considered for a monochromatic wavelength, for an all-mirror optical system wave-front at other wavelength can be simply scaled according to the wavelength. The precision of the first-order approximation can be further improved by applying an iteration approach that will be discussed in our future publications.

1    Ren D Q, Zhu Y T. A Coronagraph based on stepped-transmission filters. PASP, 2007, 119: 1063－1068
2    Guyon O, Pluzhnik E A, Kuchner M J, et al. Theoretical limits on extrasolar terrestrial planet detection with coronagraphs. ApJ，2006, 167: 81－89
3    Ren D Q, Serabyn E. Symmetric nulling coronagraph based on a rotational shearing interferometer. Appl Opt, 2005, 44: 7070－7073
4    Ren D Q and Wang H M. Spectral subtraction: a new approach to remove Low and high-order speckle noise, ApJ, 2006, 640:530–537
5    Malbet F, Yu J W, Shao M. High-dynamic range imaging using a deformable mirror for space coronagraphy. PASP,1995, 107: 386–398
6    Borde P J, Traub W A. High-contrast imaging from space: speckle nulling in a low-aberration regime. ApJ, 2006, 638: 488–498
7    Give'on A, Beliko R, Shklan S , et al. Closed loop, DM diversity-based, wavefront correction algorithm for high contrast imaging systems. Opt. Express, 2007, 15:12338－12343
8    Enya K, Tanaka S, Abe L, et al. Laboratory experiment of checkerboard pupil mask coronagraph. A&A, 2007, 461: 783–787
9    Tyson R K. Principle of Adaptive Optics. USA: Academic Press,1998